\documentclass[10pt,journal,compsoc]{IEEEtran}
\ifCLASSINFOpdf
\else
\fi
\usepackage{lipsum}
\usepackage{mathtools}
\usepackage{cuted}
\usepackage{amsmath}
\usepackage{url}
\usepackage{multirow}
\usepackage{flushend}
\usepackage{tabularx}
\usepackage{amsmath}

\begin{document}
%
\title{Experimenting with Energy-Awareness in Edge-Cloud Containerized Application Orchestration

}
%
%
%

\author{Dalal Ali,
        Rute C. Sofia,~\IEEEmembership{Senior Member, IEEE}
       \thanks{D. Ali and R. C. Sofia are with the IIoT competence field of the fortiss research institute, Munich, Germany e-mail: {ali,sofia@fortiss.org.}}}

\markboth{Pre-print, 2025}%
{D. Ali, R. C. Sofia \MakeLowercase{\textit{Ali et al.}}: Experimenting with Energy-awareness in Edge-Cloud Containerized Application Orchestration}
%



\maketitle

\begin{abstract}
This paper explores the role of energy-awareness strategies into the deployment of applications across heterogeneous Edge-Cloud infrastructures. It proposes methods to inject into existing scheduling approaches energy metrics at a computational and network level, to optimize resource allocation and reduce energy consumption. 
The proposed approach is experimentally evaluated using a real-world testbed based on ARM devices, comparing energy consumption and workload distribution against standard Kubernetes scheduling. Results demonstrate consistent improvements in energy efficiency, particularly under high-load scenarios, highlighting the potential of incorporating energy-awareness into orchestration processes for more sustainable cloud-native computing.
\end{abstract}

\begin{IEEEkeywords}
Kubernetes, Edge-Cloud continuum, resource management, energy awareness.
\end{IEEEkeywords}

%
\IEEEpeerreviewmaketitle

\section{Introduction}
\label{intro}

\IEEEPARstart{I}{nternet} of Things (IoT) applications are today deployed across the so-called Edge-Cloud continuum. These applications are composed of micro-services which can be virtualized to increase their portability and lightweight. Container technologies such as Docker are the most relied solutions to deploy IoT applications across the so-called Edge-Cloud continuum. Application management (setup, runtime) is handled by containerized application orchestrators, such as \textit{Kubernetes (K8s)}. 
Tools such as K8s play a key role in the application lifecycle management, supporting application needs in terms of scalability and availability of the required resources, and load balancing. However, originally developed for Cloud environments, such tools consider that the Edge-Cloud infrastructure relates only with computational resources, such as processing (e.g., CPU, GPU) and memory usage.

However, the overall Edge-Cloud infrastructure is composed of resources at a network, computational and data level. When considering the deployment of applications and of their micro-services, the orchestration of resources from these different perspectives is orchestrated via different solutions. For instance, at a networking level, the orchestration is handled via \textit{Software Defined Networking (SDN)} solutions. At a computational level, K8s is used to handle the orchestration of the application at a computational perspective.

Orchestration should, however, be developed in an integrated way to reduce the probability of occurrence of errors. This is the approach followed by the \textit{Cognitive Decentralised Edge-Cloud Orchestration (CODECO)} Kubernetes framework~\cite{sofia2024framework}\footnote{https://he-codeco.eu}. CODECO extends Kubernetes with a data-compute-network orchestration approach. It aims at orchestrating the application lifecycle across Edge-Cloud in a way that is agnostic to the user and yet related with performance targets that the user sets, e.g., keep a specific level of \textit{greenness} or \textit{resilience}.
CODECO aims at keeping such levels while ensuring that Quality of Service (QoS) and Quality of Experience (QoE) application requirements are met. 

This paper explores the CODECO notion of greenness in the orchestration of applications as an example on how energy-awareness can be formulated and integrated into containerized application orchestration across Edge-Cloud.

This work addresses this gap by incorporating energy-awareness into the orchestration layer considering the directions being taken in the CODECO (Cognitive Decentralized Edge-Cloud Orchestration) framework\cite{sofia2024framework, sofia2023dynamic}. This work includes the following contributions:
\begin{itemize}
    \item A multi-layer proposal to bring energy-awareness into Kubernetes scheduling approaches.
    \item A formulation on how energy metrics at a network and computational level can be combined to bring a global perspective on energy cost.
    \item Experimental validation against the plain Kubernetes showing energy consumption improvement in a realistic testbed based on far Edge (embedded) devices.
\end{itemize}

This paper is organized as follows. Section \ref{relatedwork} presents related work to ours, highlighting our key contributions. Section \ref{codeco} introduces the K8s CODECO framework, specifically focusing on the components that address the injection of context-awareness into the scheduling mechanism.  Section \ref{greenness} debates on the proposed formulation of energy awareness considering how metrics at a network and computational level may be combined to provide a global definition of energy awareness. Section \ref{Experimentation} provides the experimental settings that the paper considers for evaluating the notions of greenness. Section \ref{conclusions} concludes the paper, summarising key findings and proposing directions for future research.

\section{Related Work}
\label{relatedwork}
Rao et al propose to address the energy-awareness gap in Kubernetes via a \textit{Service Level Agreement (SLA) } oriented scheduling approach~\cite{rao2025energy}. The proposed algorithm is based on the \textit{Sparrow Search Algorithm (SSA)} to optimize Pod packing by balancing inter-Pod communication frequency and microservice resource consumption, ensuring SLA compliance while minimizing energy usage. In comparison, our work focuses on the methods to bring a cross-layer, energy-aware perspective in real-time to the scheduling process of Kubernetes.

Ghafouri et al. propose Smart-Kube which relies on \textit{Reinforced Learning (RL)} to ensure that the application QoS requirements are met, while ensuring balance in terms of energy efficiency~\cite{ghafouri2023smart}. Smart-Kube learns continuously and dynamically adjusts scheduling decisions to reduce energy consumption while maintaining equitable resource distribution. Similarly, the approach being devised in CODECO dynamically learns and adjusts scheduling decisions (rf. to section 3). While this work can be combined with RL approaches such as Smart-Kube, and while CODECO employs RL to provide recommendations to the CODECO scheduler, thus increasing fairness and resilience of the overall process, it brings the possibility to explore more complex context-aware patterns to enrich and improve the application workload deployment processes (scheduling and re-scheduling).
KEIDS is an approach by Kaud et al. which formulates task scheduling as a multi-objective integer linear programming (ILP) problem to minimize energy usage and application interference in IIoT edge-cloud environments. It ensures faster, energy-efficient scheduling with optimal application performance for end-users. Our approach is aligned with the CODECO scheduling approach (Seamless Workload Migration), which also relies on ILP to deploy applications considering context-awareness (data, compute, network) and application QoS requirements. In this work, we address only the formulation of energy-awareness and explain how it can be applied to improve the Kubernetes plain scheduling~\cite{kaur2019keids}. 
Centofanti et al. provide an extensive overview on energy monitoring approaches, proposing a comparison of different approaches in the context of Cloud compputing, such as Kepler, Scaphandre from a computational perspective~\cite{centofanti2024impact}.  The study emphasizes the importance of accurate power monitoring to enable cloud providers to deploy applications strategically, leading to more energy-efficient operations. Our work focuses on approaches that can bring energy-awareness from a cross layer perspective, to bring energy-awareness by design to the overall application orchestration mechanisms.

\section{The CODECO Framework and its Capabilities}
\label{codeco}

\begin{figure*}[!ht]
\centering
\includegraphics[width=\textwidth]{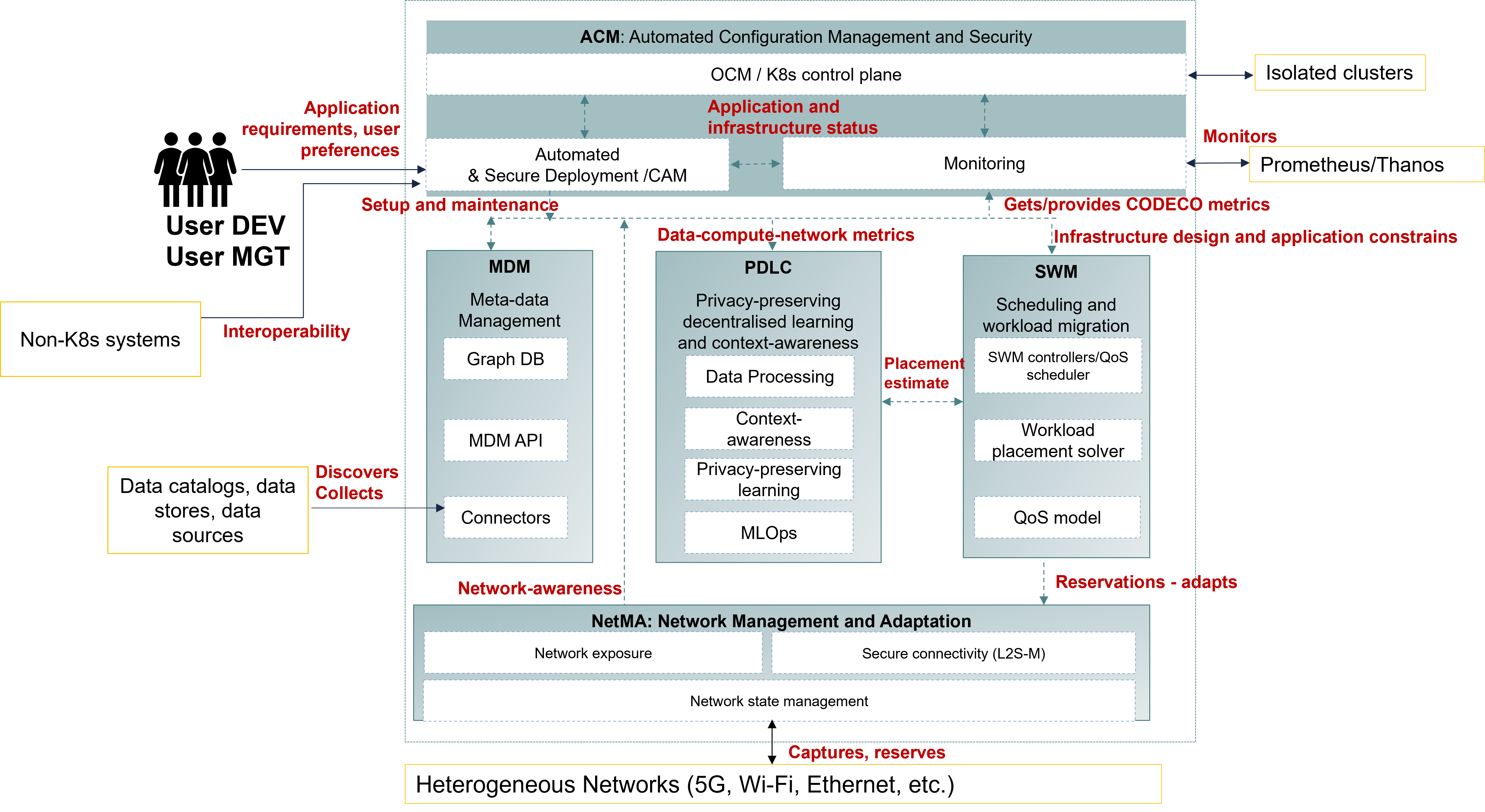}
\caption{The CODECO framework and its components.}
\label{fig:codeco}
\end{figure*}

CODECO is a Kubernetes-based framework designed to enable intelligent and flexible orchestration of applications across heterogeneous and mobile IoT-Edge-Cloud environments. It manages the full application lifecycle—from setup to runtime—by taking into account networking resources, data dependencies, and computational capabilities to optimize both deployment and execution. To achieve this, CODECO integrates data-compute-network orchestration, enabling adaptive, efficient, and intelligent resource allocation across the Edge-Cloud continuum. A functional overview of CODECO and its core components is illustrated in Figure \ref{fig:codeco}.

 CODECO enables automated application setup and runtime across Edge-Cloud environments via the \textit{Advanced Configuration and Management (ACM)} component, considering compute, network, and data aspects. The \textit{Metadata Manager (MDM)} treats data as a resource and offers multi-perspective infrastructure snapshots to detect changes.

The \textit{Scheduling and Workload Migration (SWM)} component uses QoS models for dynamic scheduling and migration, aligning applications with optimal infrastructure across clusters. The \textit{Privacy-preserving Decentralized Learning (PDLC)} component enables context-aware, real-time adjustments to processing, compute, and network resources.

The \textit{Network Management and Adaptation (NetMA)} component supports infrastructure adaptation by optimizing networking resources across clusters, from Edge to Cloud.

The only user interface in CODECO is the ACM component, which manages application setup and runtime from Edge to Cloud based on user input. Users are seen as either application developers (DEV) or cluster managers (MGT). Through the CODECO Application Model (CAM), users define application and micro-service requirements for efficient CEI deployment.  Via CAM, the user specifies also target performance profiles that can be used by the component CODECO (PDLC) to meet a specific level of performance, for instance, greenness or resilience during its operation.

To enable flexible orchestration, CODECO monitors the following infrastructure layers:  networking (via the NetMA component), data workflow and status (via the MDM component), compute (via ACM). All metrics are collected and exported to the de facto Kubernetes monitoring server, Prometheus\footnote{\url{https://prometheus.io/}}. This approach supports integration of new metrics, including the support for user-defined metrics\cite{CSofia2023DynamicCC}\footnote{A full list of infrastructure metrics monitored in CODECO is available in its Deliverable D10~\cite{D10}.}: 

\begin{itemize}
    \item \textbf{Networking Perspective:} CODECO treats networking as both an underlay and overlay infrastructure component, ensuring robust interconnectivity across Cloud and Edge environments.
    \item \textbf{Data Observability Perspective:} recognizing data dependencies, CODECO incorporates observability mechanisms that facilitate informed deployment and re-deployment decisions.
    \item \textbf{Computational Perspective:} CODECO optimizes node selection by evaluating computational resources, thereby ensuring optimal placement of microservices based on application-specific demands.
\end{itemize}

\subsection{Privacy-preserving Decentralized Learning and Context-awareness}
\begin{figure*}[!htp]
\centering
\includegraphics[width=\textwidth]{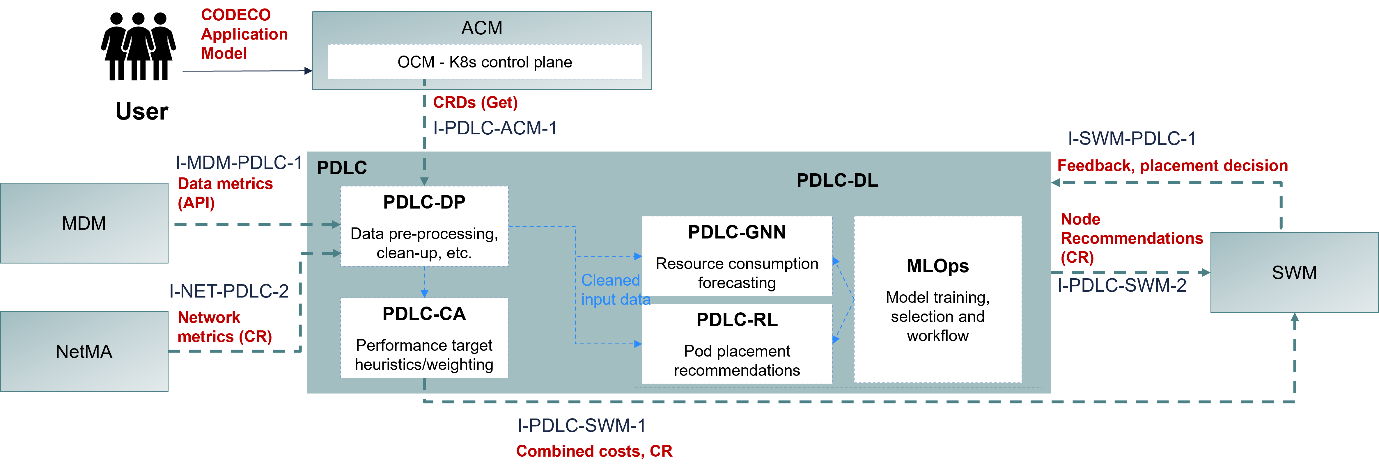}
\caption{The PDLC component in CODECO, having as input metrics available in Prometheus and collected via ACM, NetMA and MDM, and as output recommendations to the CODECO scheduler, SWM. PDLC-CA as sub-component is responsible for combining metrics based on a specified user target profile.}
\label{fig:PDLC}
\end{figure*}

PDLC as a component of CODECO plays a key role in terms of bringing context-awareness and inference to the overall orchestration. PDLC has two main roles: 

\begin{itemize}
    \item \textbf{Aggregated Node Cost Estimation:} its sub-component Context-awareness (PDLC-CA) provides aggregated node costs targeting the user-defined performance profiles, ensuring customized deployment decisions with lower complexity.
    \item \textbf{System Stability Estimation:} Using decentralized learning techniques, PDLC passes recommendations to the CODECO scheduler in terms of infrastructure that may bring more stability into the overall system, eventually contributing to resilience and system stability.
\end{itemize}
A functional representation of PDLC is provided in Figure \ref{fig:PDLC}, where the different components can be observed. As represented, CODECO introduces a strict separation between node cost recommendations provided by PDLC and the specific decision on where and when to schedule/re-schedule application micro-services (application workloads) by SWM.  This modular approach has been devised to ensure that the main scheduler in the project is the ILP-based SWM scheduler, rather than relying on PDLC components such as decentralized learning techniques as the primary scheduler. Instead, PDLC functions solely as a recommendation system to support the main scheduler. This design choice provides greater flexibility and removes the need to retrain the scheduler when new constraints or performance goals are introduced. Furthermore provides also a higher degree of interoperability, allowing CODECO to be used with different Kubernetes-aware schedulers.

\subsection{Bringing Context-Awareness to Edge-Cloud Orchestration: PDLC-CA}
PDLC-CA is a core sub-component of PDLC, responsible for creating
node aggregated costs aligned with user-defined performance profiles, such as resilience or greenness. PDLC-CA receives metrics monitored by different components of CODECO which are available via the monitoring server Prometheus, and performs metric aggregation based on proposed aggregated cost functions. The result is a node cost which provides the basis to score nodes in terms of their greenness, resilience, or other performance profiles provided by the user. Such scoring is then provided to PDLC-DL, which relies on the aggregated node costs along with infrastructure metrics to derive node costs.
PDLC-CA can also provide its output directly to Kubernetes schedulers, such as the CODECO SWM scheduler. In doing so, it is feasible to bring to K8s resilience-awareness or energy-awareness.

The next section addresses how CODECO currently explores flexible formulations of greenness and energy-efficiency.

\section{Greenness and Energy-efficiency in CODECO}
\label{greenness}

\subsection{CODECO Greenness Metrics}
To enable energy-efficient orchestration of containerized applications, CODECO considers that Greenness is a flexible concept which can be aligned with energy-efficiency, with CO2 footprinting, etc.
In this specific work, greenness is being defined as energy-awareness and as such, greenness relates with energy consumption metrics that are collected at a computational and networking level.

Currently, the following energy related metrics are considered in CODECO:

\begin{itemize}
    \item \textbf{Node Energy ($n_e(i)$)}: energy consumed by node $i$, related with all its active processes.
    \item \textbf{Link Energy ($l_e(i,j$)}: The overall energy consumed by node $i$ due to transmission of bits across its egress link $j$.
    \item \textbf{Network Energy} ($L_e(i)$): the sum of the energy consumed across all links of the source node $i$, i.e., $ L_e(i)=\sum_{j}l_e(i,j)$.
  \end{itemize}

\subsection{Cost Combination Strategies for Ranking Greenness}

CODECO proposes to employ cost functions to rank suitable nodes in terms of greenness. To capture the \textit{greenness} $g(i)$ of a node $i$, different heuristics can be considered, for instance:

\begin{align}
    g(i) &= n_e(i)  \label{eq:greenness_compute}\\
    g(i) &= nLe(i) \cdot  L_e(i) \label{eq:greenness_link}
\end{align}

Equation~\ref{eq:greenness_compute} captures a compute-centric view of greenness, focusing solely on the energy consumed by a node. Equation~\ref{eq:greenness_link} extends this by incorporating network transmission energy consumption awareness, thus being sensitive to the number and energy cost of active links.

\section{Experimental Evaluation of Greenness}
\label{Experimentation}

\subsection{Setup and Methodology}
This section outlines the experimental setup used to evaluate the use of PDLC-CA and to assess the validity of a simple heuristic for greenness when compared with the regular K8s scheduling process.

In this experimentation, PDLC-CA has been used in its standalone mode\footnote{https://shorturl.at/66lDg} to provide a proposal for energy-based scheduling, derived from the CODECO greenness formulation presented in Eq. \ref{eq:greenness_compute}.
\begin{table}[htp!]
\centering
\caption{Testbed equipment features.}
\label{tab:set-up}
\begin{tabularx}{\columnwidth}{|X|X|X|X|}
\hline
\textbf{Role} & \textbf{Device} & \textbf{Specification} & \textbf{Node identifier} \\ \hline
Master Node & Laptop & 16Gi, 8 cores & master \\ \hline
Worker Node & RPI4 & 3.8Gi, 4 cores & N1 \\ \hline
Worker Node & RPI4 & 3.8Gi, 4 cores & N2 \\ \hline
Worker Node & RPI4 & 3.8Gi, 4 cores & N3 \\ \hline
\end{tabularx}
\end{table}

\begin{figure}
    \centering
    \includegraphics[width=0.5\textwidth]{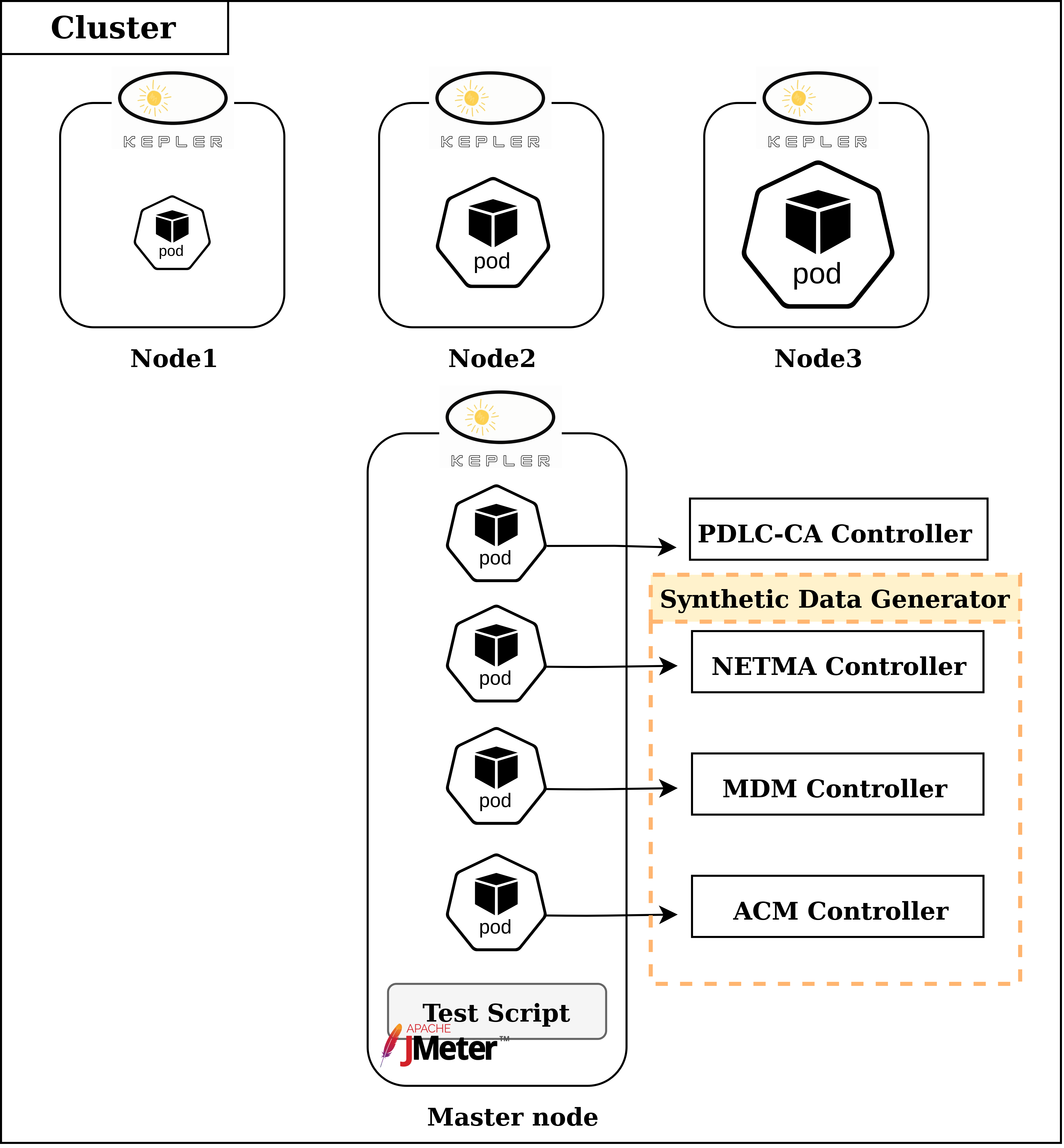}
        \caption{Experimental setup, showing the PDLC-CA and CODAG positioning in the testbed.}
    \label{fig:experimentation}
\end{figure}

The experimental environment is based on a light version of K8s distribution, K3s\footnote{https://k3s.io/} deployed on a live testbed~\footnote{https://www.fortiss.org/en/research/fortiss-labs/detail/iiot-lab}. The testbed considers a single K3s cluster integrating one master node (a laptop) and three worker nodes (Raspberry Pi 4 devices). These nodes were configured as outlined in Table \ref{tab:set-up} and interconnected via Wi-Fi (IEEE 802.11bg). The OS is Ubuntu, and the master deploys kernel version 6.8.0-57-generic while the Raspberry Pi 4 devices (RPI4) deploy kernel 6.6.22-codeco-v8+, i.e., a custom-built Kernel version considering eBPF integration due to Kepler (rf. to section 5.2).  Each node has the \textit{Kubernetes-based Efficient Power Level Exporter (Kepler)} Prometheus plugin active. Kepler relies on the \textit{extended Berkeley Packet Filter (eBPF)} to probe CPU performance counters and Linux kernel tracepoints ~\cite{kepler}. Kepler has been deployed as a \textit{daemonset} pod, responsible for monitoring the CODECO energy metrics. In CODECO, the Node Energy metric is collected by the component ACM, while the Network Energy metric is collected by the NetMA component. The metrics are then available continuously to PDLC (and to other CODECO components).
In this work, the input of PDLC-CA is provided by the \textit{CODECO Data Generator (CODAG)}\footnote{https://shorturl.at/TvyS7}, which emulates the CODECO data collection workflow\footnote{During the course of this work it was not feasible to use the full CODECO framework, including its scheduler, due to an early stage of code on some components. Due to this, we have opted to consider CODAG which emulates the monitoring process of CODECO.}. To align with our experimental requirements, the node energy queries were collected from the CODAG emulated ACM component controller which quires these metric from Prometheus and exposes them in a K8s Custom Resource Definition (CRD) for PDLC-CA to use, the same process happens with the other components to expose their metrics via CRDs for PDLC to collect them,  to aggregate them in heuristic for nodes cost which we do consider at this experiment. Figure \ref{fig:experimentation} illustrates the experimental setup, showcasing the pre-loaded configurations and the overall architecture of the testing environment.

\subsection{Kepler Energy Monitoring}
It is important to note that the experimentation cluster is based on embedded devices (ARM), and that Kepler relies on eBPF. In RPIs the \textit{BPF Type Format (BTF)} is a minimalistic, compact format, inspired by Sun's CTF (Compact C Type Format), which is used for representing kernel debug information since Solaris 9. BTF was created to allow access to hardware with a focus on simplicity and compactness to allow its usage in the Linux kernel. Access to energy metrics in ARM-based devices is not supported via the usual interfacing such as ACPI, IPMI, or Redfish \cite{RPIs}. As a result, real-time energy metrics cannot be collected directly based on hard-sensing. In contrast, in x86-based bare-metal environments with internal sensors, Kepler can directly access power data from the kernel and export it at both node and container levels using eBPF as described in Kepler’s official documentation \cite{kepler}.

Kepler’s pre-trained power models, based on CPU and system resource utilization, provide general power estimations but lack the precision of real-time metrics\cite{cncf2023}. These models enable the estimation of dynamic energy consumption but are architecture-specific and trained on reference hardware (e.g., Intel\textsuperscript{\textregistered} Xeon\textsuperscript{\textregistered} E5-2667 v3). Although convenient, they can produce inaccurate results when deployed on mismatched hardware. 

This work considers Kepler’s default x86-based model to estimate energy consumption on ARM-based nodes. This provides a general approximation of energy consumption behavior, but we acknowledge its limitations. For improved accuracy, architecture-specific model training \cite{RPIs} is preferable. When retraining is not feasible, empirical methods such as the average wattage estimation proposed by Pol et al.  offer an alternative \cite{Pol2024}.

The Node Energy metric (rf. to Eq. 1) has been collected via the Kepler query represented in (\ref{eq:keplerquery}). The query provides a cumulative perspective on the energy consumed every five minutes.

\begin{equation}
\begin{aligned}
\texttt{query} = \texttt{increase}(&\texttt{kepler\_node\_platform} \\ &\texttt{\_joules\_total\{exported} \\ 
&\texttt{\_instance="<node.name>"}, \\
&\texttt{mode="dynamic"}\texttt{\}[5m]})
\end{aligned}
\label{eq:keplerquery}
\end{equation}

\subsection{Scenario}
Experiments rely on Apache JMeter \cite{apache_jmeter} with a simple Flask-based Python application that performs matrix multiplication. Each request generates two random matrices (with dimensions up to 3×3) and computes their product. Two primary test scenarios are defined based on the request rate:

\begin{itemize}
\item \textbf{Moderate Load Intensity:} 5, 10, and 15 \textit{requests per second (rps)}
\item \textbf{High Load Intensity:} 15, 20, and 25 rps
\end{itemize}

Request modeling in JMeter involves configuring both the request rate and the concurrency level. The \textit{Concurrency Thread Group}\footnote{\url{https://jmeter-plugins.org/wiki/ConcurrencyThreadGroup/}} maintains a consistent number of active threads to sustain the target load. The required number of threads $C$ is calculated as:

\begin{equation}
C = \frac{r \cdot t}{1000}
\label{Eq-thread}
\end{equation}

where $r$ is the target request rate (requests per second), and $t$ is the expected maximum response time in milliseconds (here, 250ms). This ensures that threads remain active long enough to sustain continuous pressure on the system, supporting stable energy and performance measurements. The JMeter \textit{Throughput Shaping Timer} controls the pacing of requests across all runs.

Experiments considered a warm-up time of 5 minutes to ensure metric collection consistency, while a 8-minute cool-down period between experiments was used to ensure that results were not influenced by prior runs. 

To assess energy consumption across different deployment patterns, three main configurations were tested:

\begin{itemize}
\item \textbf{Configuration 1:} No pre-loaded pods (baseline configuration).
\item \textbf{Configuration 2:} Uniform load distribution — each worker node hosted three pods (10 pods in the case of moderate load and 20 pods for high load scenarios).
\item \textbf{Configuration 3:} Heterogeneous load distribution across the worker nodes:
  \begin{itemize}
  \item \textbf{3A:} High demand on node \texttt{N1}, medium demand on \texttt{N2}, and low demand on \texttt{N3}.
  \item \textbf{3B:} Low demand on node \texttt{N1}, high demand on \texttt{N2}, and medium demand on \texttt{N3}.
  \item \textbf{3C:} Medium demand on node \texttt{N1}, low demand on \texttt{N2}, and high demand on \texttt{N3}.
  \end{itemize}
\end{itemize}

In all configurations, the test pod, scheduled after the setup of each configuration as seen in the breakdown, consistently received medium demand relative to to the load intensity set, regardless of the configuration variation.

Table \ref{tab:config_summary} summarizes the experimental configurations, including request rates and pod sizes. Pod sizes reflect different CPU resource allocations (in millicores), representing varying processing capacities across nodes. The application choice, along with the light matrix dimensions (up to 3×3), response time (250 ms), and CPU allocations (300–500 millicores), are exploratory and not based on detailed application profiling or tuning. At this initial stage of the study, these parameters were selected to generate representative load and energy consumption patterns, as the primary goal is to analyze energy behavior rather than optimize performance.
\begin{table}[htp!]
\caption{Summary of experimental configuration (S = Small impact in terms of frequency requests (300 millicore, 5 or 15 RPS), M = Medium (400 millicore, 10 or 20 RPS), L = Higher impact in requests (500 millicore, 15 or 25 RPS)).}
\small
\label{tab:config_summary}
\resizebox{\columnwidth}{!}{%

\setlength{\tabcolsep}{2pt}
\begin{tabular}{|p{2cm}|p{5cm}|p{4cm}|}
\hline
\textbf{Configuration} & \textbf{Nodes (Pod Size, Load RPS)} & \textbf{Setup} \\ \hline
\textbf{1} & All Nodes (None) & Baseline, no preloaded pods \\ \hline
\textbf{2} & All Nodes [M] (10 or 20 RPS) & Uniform setup \\ \hline

\textbf{3A} & N1 [L] (15 or 25 RPS) & \multirow{3}{*}{Varying sizes and loads} \\ \cline{2-2}
           & N2 [M] (10 or 20 RPS) & \\ \cline{2-2}
           & N3 [S] (5 or 15 RPS) & \\ \hline

\textbf{3B} & N1 [S] (5 or 15 RPS) & \multirow{3}{*}{Varying sizes and loads} \\ \cline{2-2}
           & N2 [L] (15 or 25 RPS) & \\ \cline{2-2}
           & N3 [M] (10 or 20 RPS) & \\ \hline

\textbf{3C} & N1 [M] (10 or 20 RPS) & \multirow{3}{*}{Varying sizes and loads} \\ \cline{2-2}
           & N2 [S] (5 or 15 RPS) & \\ \cline{2-2}
           & N3 [L] (15 or 25 RPS) & \\ \hline

\end{tabular}}
\end{table}

\begin{figure*}[h!]
\centering
 \fbox{\includegraphics[width=0.8\textwidth]{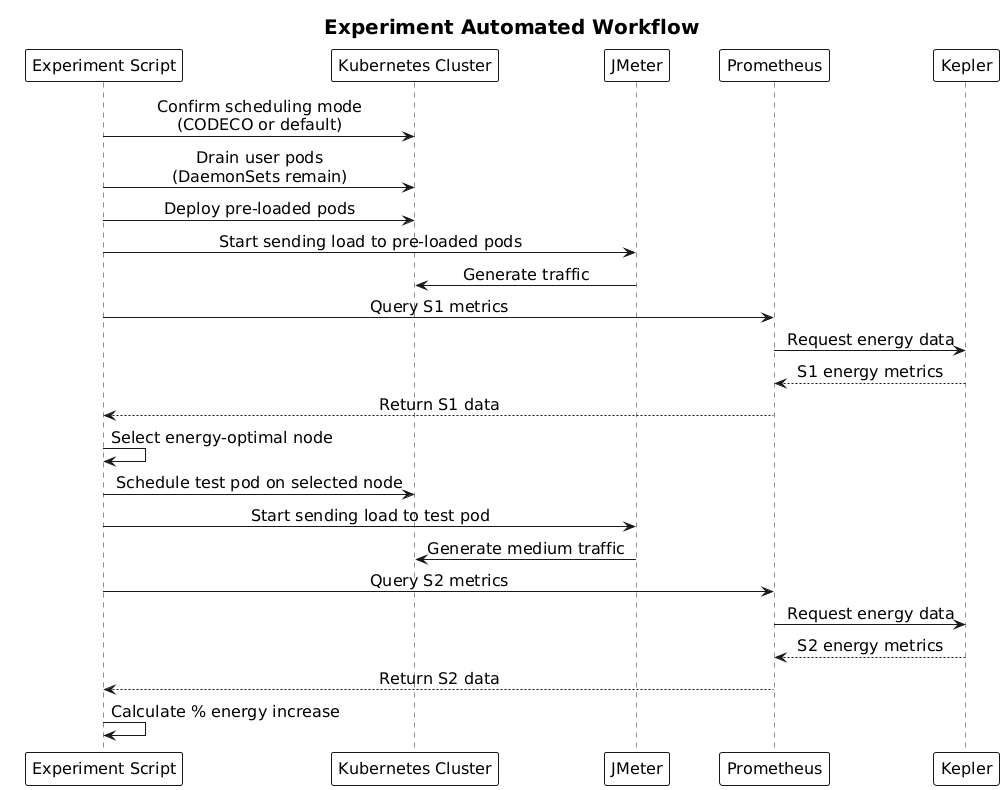}}
\caption{Communication sequence for the overall experimental workflow described in this paper.}
\label{fig:workflow}
\end{figure*}

\subsection{Experiment Automated Workflow}

Experiments were automated via a python script to ensure reproducibility across configurations as represented  in Figure~\ref{fig:workflow}. 
The script executes the following steps:

\begin{itemize}
    \item Confirms the active scheduling mode (\texttt{CODECO} scheduling strategy or the default Kubernetes scheduling strategy).
    
    \item Prepares the cluster environment by draining all worker nodes of user-deployed pods. DaemonSet pods, such as those for Kepler and K3s system services, remain active, as they are consistently present on each node.
    
    \item Deploys required pre-loaded service pods for configurations that demand them.
    
    \item Uses Apache JMeter to simulate a baseline workload across the cluster.
    
    \item Collects dynamic energy metrics during this first phase (S1) using Kepler via Prometheus queries in emulated ACM controller which is exposed as a Custom Resource (CR) to PDLC-CA. Based on the measured values, the script determines the most energy-efficient node.
    
    \item Schedules the target test pod onto the selected energy-optimal node.
    
    \item Simulates the medium demand load directed at the test pod using JMeter.
    
    \item Collects energy metrics during this second phase (S2) and calculates the percentage increase in energy consumption between S1 and S2 for both scheduling strategies.
\end{itemize}

\section{Performance Evaluation and Analysis}
The performance evaluation addresses the impact of the CODECO energy-aware scheduling strategy against the plain  Kubernetes scheduler. For this purpose, a measure of such impact is $ei$ formulated in Eq. \ref{eq:ei}. $ei$ corresponds to the relative increase of energy consumed in the cluster from the instant when the application is going to be deployed, $e_{s1}$ (\textit{S1, initial setup}), until after the application has been deployed, $e_{s2}$ (\textit{S2, post-deployment}). 

\begin{equation}
\begin{aligned}
ei=\frac{e_{s2}-e_{s1}}{e_{s1}}\cdot 100
\end{aligned}
\label{eq:ei}
\end{equation}

\subsection{Moderate Load Analysis}
A first batch of experiments has been run considering moderate load configurations. Results are presented in Table \ref{tab:low} for each application deployment stage (S1, S2). The table compares the energy consumption performance for a single cluster with the CODECO energy-aware optimized orchestration approach, and the baseline K8s orchestration. The metrics considered are:
\begin{itemize}
    \item Total cluster energy consumption in Joules for each of the stages, $e_c(S1)$ and $e_c(S2)$
    \item The relative increase in energy consumption from S1 to S2, $e_i$
    \item The node selected (Nx) and number of runs where the node was selected (y).
\end{itemize}

The total cluster energy has been computed for 5 runs and averaged out to get an estimate of a representative total cluster energy. The total cluster energy increase between the two stages is also presented, followed by the selected nodes by K8s.

In \textit{Config 1}, where the test pod was isolated, it can first be observed that there is a decrease in the energy consumed when the CODECO greenness approach is applied to the scheduling. Another observation that can be seen is that there is a significant increase in the overall energy from S1 to S2, as expected. This is due to the fact that, while the vanilla Kubernetes scheduler selects nodes based on a scoring mechanism, favoring those with the most available resources, the CODECO greenness scheduling provides a selection based on nodes that are using less energy for their resources. Overall, in this setting, CODECO consumes less total energy post-deployment and shows a lower energy increase than K8s. Its node selection favors N3 more times, showing slightly more consistency than K8s.

In \textit{Config 2}, the K8s approach brings again slightly higher results in terms of energy consumption. This can also be seen in the selection of nodes by both scheduling approaches. Node \texttt{N1} was the node that was most selected in the different runs for K8s scheduling, while for the CODECO greenness approach, node \texttt{N3} (chosen 4 times) was the preferred one. 

\textit{Configs 3.1, 3.2, 3.3 }are similar in the sense that they bring different loads (small, medium, large) to specific nodes, being the different the number of requests per second. For these specific configurations, it is relevant to highlight that while in S1 the energy consumed is always lower for the CODECO greenness scheduling, that is not the case for the case of S2, where Configs 3.1 and 3.3 result in slightly more energy being used with the CODECO greenness scheduling. It is also observable that the improvement seen in the CODECO greenness scheduling is based on the selection of the same node that the K8s scheduler has considered and corresponds to the worker node with the smallest workload configuration. 

\begin{table}[htp!]
\centering
\caption{Impact on energy consumption for scenarios with moderated load of requests.\label{tab:low}}
\small
\resizebox{\columnwidth}{!}{%
\begin{tabular}{|p{1cm}|p{3cm}|p{3cm}|p{3cm}|}
\hline
\textbf{Set} & \textbf{Metric} & \textbf{CODECO} & \textbf{Kubernetes} \\
\hline

\multirow{4}{*}{1}
& $e_{s1}$ & 4.963 ($\pm$ 1.021) & 5.0174 ($\pm$ 0.772) \\
& $e_{s2}$ & 75.311 ($\pm$ 2.535) & 91.316 ($\pm$ 11.389) \\
& $ei$     & 93.410\% & 94.505\% \\
& Chosen Node & N3 (4), N2 (1) & N1 (3), N2 (2) \\
\hline

\multirow{4}{*}{2}
& $e_{s1}$ & 242.166 ($\pm$ 4.918) & 238.968 ($\pm$ 7.747) \\
& $e_{s2}$ & 269.114 ($\pm$ 2.638) & 285.297 ($\pm$ 5.902) \\
& $ei$     & 10.014\% & 16.239\% \\
& Chosen Node & N3 (4), N2 (1) & N1 (4), N2 (1) \\
\hline

\multirow{4}{*}{3.1}
& $e_{s1}$ & 275.574 ($\pm$ 5.850) & 280.969 ($\pm$ 5.845) \\
& $e_{s2}$ & 325.750 ($\pm$ 2.860) & 324.944 ($\pm$ 11.400) \\
& $ei$     & 15.403\% & 13.533\% \\
& Chosen Node & N3 (5) & N3 (5) \\
\hline

\multirow{4}{*}{3.2}
& $e_{s1}$ & 213.339 ($\pm$ 10.382) & 218.948 ($\pm$ 7.912) \\
& $e_{s2}$ & 277.458 ($\pm$ 4.305) & 280.759 ($\pm$ 7.867) \\
& $ei$     & 27.107\% & 24.081\% \\
& Chosen Node & N1 (5) & N1 (5) \\
\hline

\multirow{4}{*}{3.3}
& $e_{s1}$ & 266.428 ($\pm$ 2.705) & 269.943 ($\pm$ 9.956) \\
& $e_{s2}$ & 326.124 ($\pm$ 9.952) & 315.759 ($\pm$ 5.818) \\
& $ei$     & 18.305\% & 14.510\% \\
& Chosen Node & N2 (5) & N2 (5) \\
\hline

\end{tabular}}
\end{table}

In Config 3.2 (medium load), CODECO maintains lower energy even after the workload is processed. However, in Configs 3.1 and 3.3, CODECO results in slightly higher energy consumption compared to K8s in S2. CODECO shows a larger energy increase from S1 to S2 in all three configurations. This indicates that under both low and high workloads, CODECO’s energy-aware scheduling may incur overhead (e.g., background monitoring, aggressive reallocation) that outweighs its static energy savings. However, it may also be due to the issues mentioned with Kepler models.
Across all configurations, CODECO selects more times N3, and seems to avoid N1, which is considered by K8s several times. Hence, CODECO avoids less efficient nodes (like N1), leading to lower total energy use. When it deviates from K8s and avoids energetically expensive nodes (notably N1), CODECO consistently outperforms in both total energy and efficiency gains. However, in cases where both strategies select the same node, CODECO does not show a clear advantage. This suggests that under moderate load conditions, background processes, monitoring daemons, and other system-level effects can introduce non-negligible energy variability, even when node choices are identical.  Moreover, while every effort was made to isolate test runs and maintain consistent conditions across strategies, some variability is inevitable when testing in real environments and without using an exact power model for the RPis. This suggests that training the energy estimation model on the precise power characteristics of the RPi nodes could further enhance CODECO’s effectiveness.

\subsection{High Load Intensity}
Table \ref{tab:high} provides the results for the same configs, now considering the models for high load of requests per second.
Overall, the CODECO greenness scheduling approach results in less energy across all cases for S1. Interestingly, it also now shows lower energy consumption of the system across S2. Hence, overall the energy consumption increase is lower for CODECO. This seems to again hint that for environments where there is a higher degree of variability (different number of RPS, different node loads), CODECO benefits are more clear.

Furthermore, what can be observed across all cases, is that CODECO seems to avoid N1 and prefer N3 which seems to be the node more energetically efficient. This may be beneficial for environments where applications and their micro-services may have to be migrated frequently, e.g, environments where nodes are energy-constrained or subject to other resource constraints, e.g., network connectivity.

In high-load scenarios, CODECO’s energy efficiency largely stems from intelligent node selection. Avoiding nodes like N1, which may have higher idle or dynamic energy costs due to system services or hardware traits, results in meaningful reductions in energy use. Even in cases where node choices are identical, CODECO occasionally demonstrates lower runtime overhead, suggesting it benefits not only from greener scheduling decisions but also from leaner orchestration execution.

When comparing the performance of both approaches in scenarios with medium load of requests and high load of requests, it can be seen that for the case of high loads, there is a consistent increase in the energy consumed in both stages, as expected. In regards to energy efficiency, high-load results show that CODECO brings a more consistent energy advantage even when node selection is the same. Under moderate loads, the difference is less pronounced, possibly due to lower energy demand or lower scheduler impact on energy dynamics. 

In regards to node selection, CODECO consistently avoids N1, indicating it recognizes N1's higher energy footprint. When node choice differs, CODECO outperforms K8s. When node choice is the same, performance is often similar or only slightly better for CODECO.

\begin{table}[htp!]
\centering
\caption{Impact on energy consumption for scenarios with higher load of requests.\label{tab:high}}
\small
\resizebox{\columnwidth}{!}{%
\begin{tabular}{|p{1cm}|p{3cm}|p{3cm}|p{3cm}|}
\hline
\textbf{Set} & \textbf{Metric} & \textbf{CODECO} & \textbf{Kubernetes} \\
\hline

\multirow{4}{*}{1}
& $e_{s1}$ & 5.556 ($\pm$ 0.501) & 5.764 ($\pm$ 0.378) \\
& $e_{s2}$ & 104.656 ($\pm$ 6.043) & 120.119 ($\pm$ 22.119) \\
& $ei$     & 94.682\% & 95.201\% \\
& Chosen Node & N3 (4) \& N2 (1) & N1 (4) \& N2 (1) \\
\hline

\multirow{4}{*}{2}
& $e_{s1}$ & 346.930 ($\pm$ 3.806) & 348.605 ($\pm$ 6.071) \\
& $e_{s2}$ & 381.445 ($\pm$ 4.935) & 395.226 ($\pm$ 13.375) \\
& $ei$     & 9.049\% & 11.796\% \\
& Chosen Node & N3 (5) & N1 (3), N2 \& N3 (1) \\
\hline

\multirow{4}{*}{3.1}
& $e_{s1}$ & 353.841 ($\pm$ 5.251) & 354.287 ($\pm$ 6.993) \\
& $e_{s2}$ & 390.780 ($\pm$ 6.018) & 390.293 ($\pm$ 3.230) \\
& $ei$     & 9.453\% & 9.225\% \\
& Chosen Node & N3 (5) & N3 (5) \\
\hline

\multirow{4}{*}{3.2}
& $e_{s1}$ & 343.501 ($\pm$ 5.354) & 348.125 ($\pm$ 4.512) \\
& $e_{s2}$ & 374.971 ($\pm$ 7.503) & 398.165 ($\pm$ 11.639) \\
& $ei$     & 8.393\% & 12.568\% \\
& Chosen Node & N3 (5) & N1 (5) \\
\hline

\multirow{4}{*}{3.3}
& $e_{s1}$ & 343.921 ($\pm$ 3.108) & 346.767 ($\pm$ 2.822) \\
& $e_{s2}$ & 376.111 ($\pm$ 4.321) & 386.849 ($\pm$ 6.436) \\
& $ei$     & 8.559\% & 10.361\% \\
& Chosen Node & N2 (5) & N2 (5) \\
\hline

\end{tabular}%
}
\end{table}

\section{Conclusions and Next Steps}
\label{conclusions}

This work presents an initial step toward energy-aware orchestration in Kubernetes-based environments through the PDLC-CA component of the CODECO framework. Experimental validation on a real-world testbed demonstrated the potential of context-driven, CODECO energy-aware strategy for scheduling.

CODECO's energy savings at high loads are more robust, likely due to its ability to optimize under stress, minimize overhead, and allocate loads more predictively. These advantages are less pronounced under moderate workloads, where intelligent scheduling benefits may be offset by system idling or uniformity.

In Stage 1, under moderate load, the higher standard deviation is mainly caused by background system activities and normal measurement noise before the workload begins. As the load increases, the impact of these background processes diminishes relative to workload energy, resulting in lower variability and more consistent energy readings.

In Stage 2, when the workload is running, CODECO consistently shows lower or more stable standard deviation values, particularly under moderate load. This suggests that CODECO’s energy-aware scheduling results in more predictable and efficient resource use. In contrast, Kubernetes shows greater variability in energy consumption, likely due to its resource-based node selection, which leads to less consistent energy profiles and container overhead. Overall, CODECO not only reduces total energy consumption but also delivers more stable and reliable performance.

Under moderate load, CODECO provides gains primarily by avoiding suboptimal nodes (like N1). However, when node selection is the same, the advantage may diminish or reverse. Under high load, CODECO consistently outperforms Kubernetes through strategic node selection and energy-efficient task execution, even with identical node usage. This indicates that CODECO is better suited for production-grade or resource-intensive environments, where energy-aware orchestration provides clearer benefits.

A key advantage of CODECO is its ability to consistently select the same optimal nodes across multiple runs and configurations with uniform node conditions (e.g., Config 1 and 2), repeatedly identifying nodes like N3.

In conclusion, while Kubernetes uses dynamic scheduling strategies such as round-robin for load distribution and availability, these approaches do not prioritize energy efficiency. In contrast, CODECO adapts to real-time energy conditions, consistently selecting the most energy-efficient nodes. This makes it a robust solution for energy-efficient scheduling, particularly in production environments where predictability, stability, and energy efficiency are critical.

The findings also emphasize the importance of incorporating multi-layer metrics and user-defined performance profiles into orchestration processes. They reveal limitations in observability tools like Kepler, especially in heterogeneous hardware environments, highlighting the need for architecture-specific, calibrated power models.

Looking ahead, further development of CODECO will focus on enhancing cost functions, supporting federated clusters, and incorporating the full CODECO scheduling stack. These advancements will enable sustainable, context-aware orchestration in next-generation Cloud-Edge-IoT computing environments.

The following areas will be addressed in our future work:
\begin{itemize}
\item \textbf{Development of More Performance Profiles}: Creating profiles that encompass a broader range of application requirements and user behaviors for improved resource management.
\item \textbf{Refinement of Cost Functions}: Enhancing adaptability to mobile environments to ensure real-time alignment with performance goals.
\item \textbf{Integration of Federated Cluster Architectures}: Addressing the complexities of mobile nodes and heterogeneous systems while incorporating user-to-cluster metrics.
\item \textbf{Utilization of Forecast Models}: Leveraging historical node metrics to predict performance profile costs more accurately, enhancing decision-making through anticipation of future performance.
\end{itemize}

\section*{Acknowledgement}
This work has been funded by The European Commission in the context of the Horizon Europe CODECO project under grant number 101092696, and by SGC, Grant agreement nr: M-0626, project SemComIIoT.


%



%

\def\refname{REFERENCES}
\bibliographystyle{ieeetr}
\bibliography{references}

@article{centofanti2024impact,
  title={Impact of power consumption in containerized clouds: A comprehensive analysis of open-source power measurement tools},
  author={Centofanti, Carlo and Santos, Jos{\'e} and Gudepu, Venkateswarlu and Kondepu, Koteswararao},
  journal={Computer Networks},
  volume={245},
  pages={110371},
  year={2024},
  publisher={Elsevier}
}

@article{kaur2019keids,
  title={KEIDS: Kubernetes-based energy and interference driven scheduler for industrial IoT in edge-cloud ecosystem},
  author={Kaur, Kuljeet and Garg, Sahil and Kaddoum, Georges and Ahmed, Syed Hassan and Atiquzzaman, Mohammed},
  journal={IEEE Internet of Things Journal},
  volume={7},
  number={5},
  pages={4228--4237},
  year={2019},
  publisher={IEEE}
}

@inproceedings{ghafouri2023smart,
  title={Smart-kube: Energy-aware and fair kubernetes job scheduler using deep reinforcement learning},
  author={Ghafouri, Saeid and Abdipoor, Sina and Doyle, Joseph},
  booktitle={2023 IEEE 8th International Conference on Smart Cloud (SmartCloud)},
  pages={154--163},
  year={2023},
  organization={IEEE}
}

@article{rao2025energy,
  title={Energy-aware Scheduling Algorithm for Microservices in Kubernetes Clouds},
  author={Rao, Wei and Li, Hongjian},
  journal={Journal of Grid Computing},
  volume={23},
  number={1},
  pages={1--22},
  year={2025},
  publisher={Springer}
}

@article{sofia2024framework,
  title={A framework for cognitive, decentralized container orchestration},
  author={Sofia, Rute C and Salomon, Josh and Ferlin-Reiter, Simone and Garc{\'e}s-Erice, Luis and Urbanetz, Peter and Mueller, Harald and Touma, Rizkallah and Espinosa, Alejandro and Contreras, Luis M and Theodorou, Vasileios and others},
  journal={IEEE Access},
  year={2024},
  publisher={IEEE}
}

@misc{apache_jmeter,
  title = {Apache JMeter: User's Manual},
  author = {{Apache Software Foundation}},
  year = {2011},
  url = {https://svn.apache.org/repos/asf/jmeter/branches/docs-2.13/docs/usermanual/intro.html}
}

@misc{kepler,
  title = {Kepler: A Sustainable Computing Framework},
  author = {{Sustainable Computing}},
  year = {2023},
  url = {https://sustainable-computing.io/design/metrics/},
  note = {Accessed: 2024-10-22}
}

@inproceedings{RPIs,
author = {Ardito, Luca and Torchiano, Marco},
title = {Creating and evaluating a software power model for linux single board computers},
year = {2018},
isbn = {9781450357326},
publisher = {Association for Computing Machinery},
address = {New York, NY, USA},
url = {https://doi.org/10.1145/3194078.3194079},
doi = {10.1145/3194078.3194079},
booktitle = {Proceedings of the 6th International Workshop on Green and Sustainable Software},
pages = {1–8},
numpages = {8},
location = {Gothenburg, Sweden},
series = {GREENS '18}
}

@misc{cncf2023,
  author       = {{Cloud Native Computing Foundation}},
  title        = {Exploring Kepler's potentials: Unveiling cloud application power consumption},
  year         = {2023},
  month        = oct,
  day          = {11},
  url          = {https://www.cncf.io/blog/2023/10/11/exploring-keplers-potentials-unveiling-cloud-application-power-consumption/},
  note         = {Accessed: 2024-11-13}
}

@article{CSofia2023DynamicCC,
  title={Dynamic, Context-Aware Cross-Layer Orchestration of Containerized Applications},
  author={Rute C. Sofia and Douglas Dykeman and Peter Urbanetz and Akram Galal and Dushyant Dave},
  journal={IEEE Access},
  year={2023},
  volume={11},
  pages={93129-93150},
  url={https://api.semanticscholar.org/CorpusID:261160007}
}

@misc{Pol2024,
  author       = {Pol, Ties},
  title        = {Carbon Footprint Monitoring up to Container-Level in Virtualized Environments: A Hardware and Hypervisor-Free Approach},
  year         = {2024},
  type         = {Master's Thesis / Essay},
  institution  = {University of Groningen},
  degree       = {Computing Science},
  supervisor   = {Andrikopoulos, V. and Setz, B.},
  url          = {https://fse.studenttheses.ub.rug.nl/id/eprint/32271},
  note         = {Accessed: 2024-04-12}
}

@article{sofia2023dynamic,
  title={Dynamic, context-aware cross-layer orchestration of containerized applications},
  author={Sofia, Rute C and Dykeman, D and Urbanetz, P and Galal, A and Dave, D},
  journal={IEEE Access},
  volume={11},
  pages={93129--93150},
  year={2023}
}

@misc{D10,
  author       = {C. Sofia (Ed.) et al., Rute},
  title        = {CODECO Deliverable D10 - Technological Guidelines,
                   Reference Architecture, and Open-source Ecosystem
                   Design
                  },
  month        = jul,
  year         = 2024,
  publisher    = {Zenodo},
  doi          = {10.5281/zenodo.12819444},
  url          = {https://doi.org/10.5281/zenodo.12819444},
}

%

\begin{IEEEbiography}
[{\includegraphics[width=1in,height=1.25in,clip,keepaspectratio]{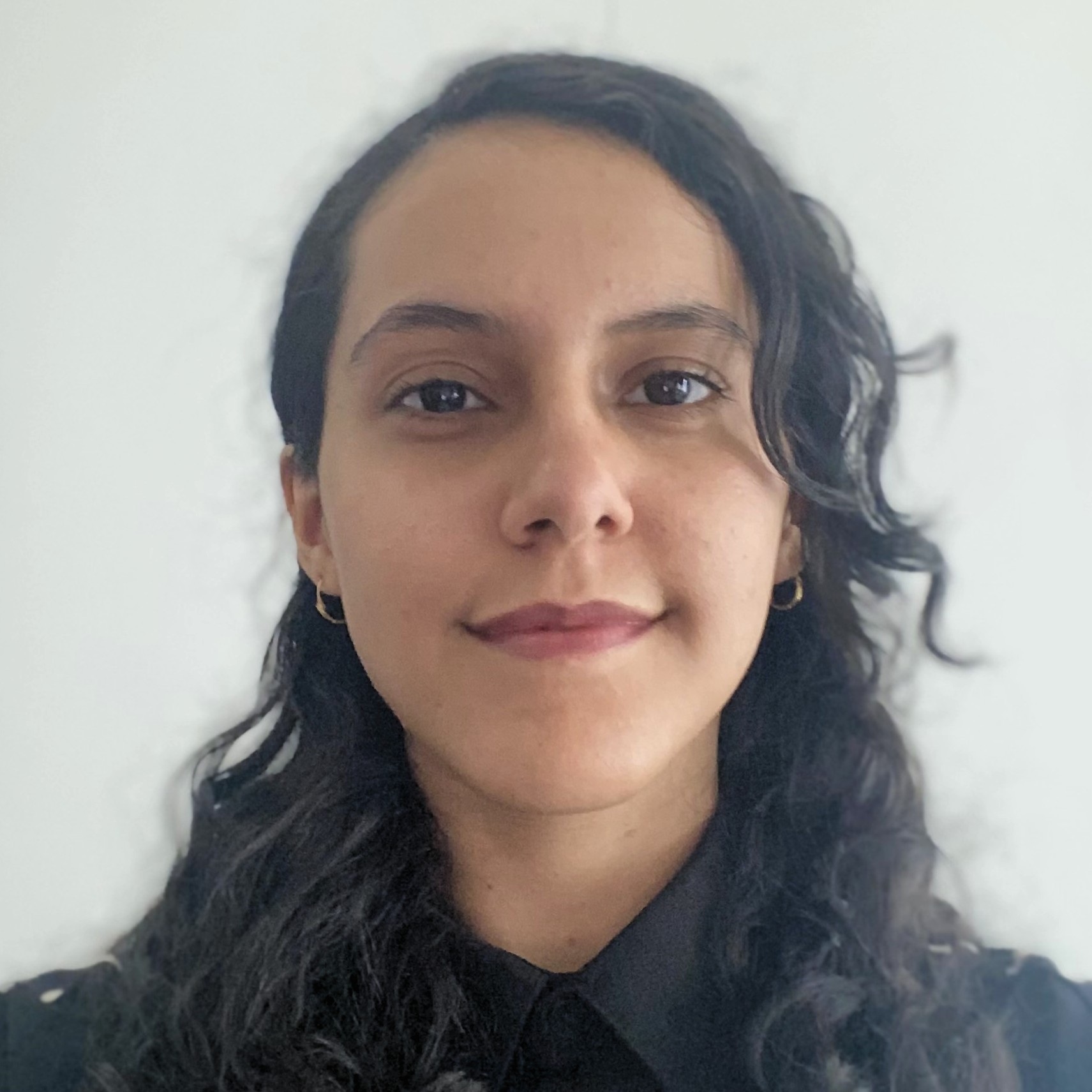}}]{Dalal Ali}
Dalal Ali is a Research Assistant in the Industrial IoT department at fortiss GmbH, Munich, and an M.Sc. Communication Engineering student at the Technical University of Munich. She contributes to the HE CODECO project, focusing on the development and performance evaluation of PDLC-CA. Her research interests include cloud and edge computing, machine learning, and the Internet of Things (IoT).
\end{IEEEbiography}


\begin{IEEEbiography}
[{\includegraphics[width=1in,height=1.25in,clip,keepaspectratio]{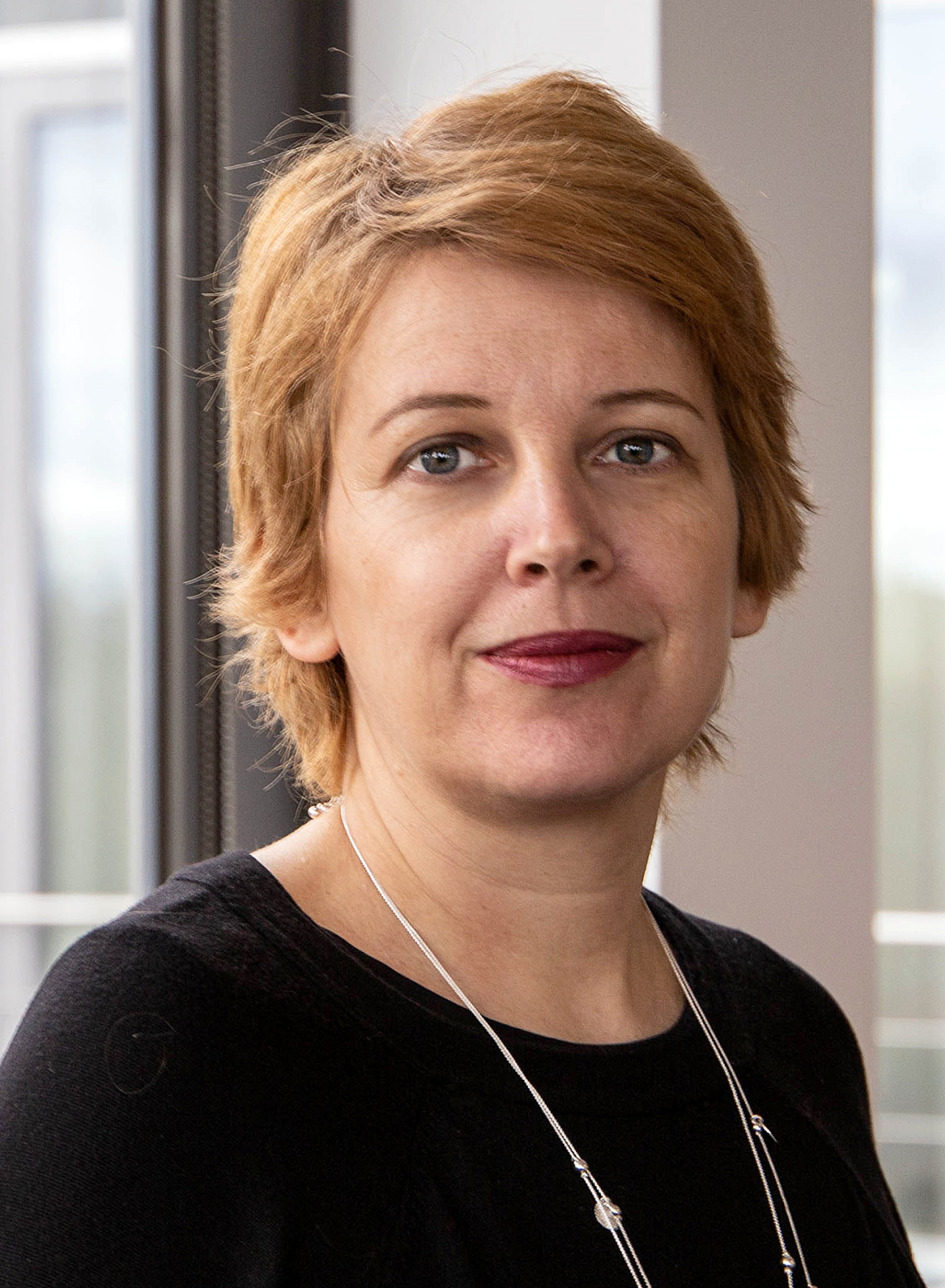}}]{Rute C. Sofia} (PhD 04, IEEE Senior Member) is the Industrial IoT department head at the research institute fortiss. She is also an Invited Associate Professor of University Lusófona de Humanidades e Tecnologias,  and an Associate Researcher at ISTAR, Instituto Universitário de Lisboa. Rute's research background has been developed on industry (Grupo Forum, Lisbon; Siemens AG, Nokia Networks, Munich) and on academia (FCCN, Lisbon; INESC TEC, Porto; ULHT, Lisbon; Bundeswehr Universität, Munich). She was a co-founder of the portuguese COPELABS research unit, and was the COPELABS scientific director (2013-2017), where she was a Senior Researcher (2010-2019). She has also co-founded the COPELABS spin-off Senception Lda (2013-2019).

Her current research interests are: network architectures and protocols; IoT; Edge computing; Edge AI; in-network computing; 6G. Rute holds over 70 peer-reviewed publications in her fields of interest, and 9 patents. She is an ACM Europe Councilor; an ACM Senior member, an IEEE Senior Member. She was an IEEE ComSoc N2Women Awards co-chair (2020-2021), and is the IEEE ComSoc WICE industry liaison deputy. She leads the 6G CONASENSE platform and the HE CODECO project. She is an associated editor among several venues, such as IEEE Network, IEEE Access.\end{IEEEbiography}


\vfill



\end{document}